\documentclass[12pt]{article}
\usepackage{amsmath,amssymb}
\usepackage{graphics}
\usepackage[dvipdfm]{graphicx}

\usepackage{cite}

\begin{document}
\baselineskip=16pt
\begin{titlepage}
\begin{flushright}
{\small OU-HET 706/2011}\\
\end{flushright}
\vspace*{1.2cm}

\begin{center}

{\Large\bf 
Spontaneous Parity Violation in SUSY Strong Gauge Theory 
} 
\lineskip .75em
\vskip 1.5cm

\normalsize
$^1${\large Naoyuki Haba}, 
and
$^2${\large Hiroshi Ohki}

\vspace{1cm}

$^1${\it Department of Physics, 
 Osaka University, Toyonaka, Osaka 560-0043, 
 Japan} \\

$^2${\it Kobayashi-Maskawa Institute for the Origin of Particles and the
 Universe, Nagoya University, Nagoya, Aichi 464-8602 Japan}

\vspace*{10mm}

{\bf Abstract}\\[5mm]
{\parbox{13cm}{\hspace{5mm}
%

We suggest simple models of spontaneous parity violation 
 in supersymmetric strong gauge theory. 
We focus on left-right symmetric model and 
 investigate vacuum with spontaneous parity violation.  
Non-perturbative effects are calculable in 
 supersymmetric gauge theory, and 
 we suggest two new models. 
The first model shows confinement, 
 and the second model has a dual description 
 of the theory. 
The left-right symmetry breaking and 
 electroweak symmetry breaking are 
 simultaneously occurred 
 with the suitable energy scale hierarchy. 
The second model also induces spontaneous 
 supersymmetry breaking.

}}

\end{center}

\end{titlepage}

\section{Introduction}

The standard model (SM) is regarded as an effective 
 theory below a TeV-scale. 
There must be a fundamental theory beyond the SM,
 and a trial of searching it is a big challenge 
 at today's experiments. 
There are some mysteries in the SM, and 
 one of them is a question, 
 ``why is the SM chiral gauge theory?''.
There is no explanation why weak interaction 
 is $SU(2)_L$, and left-right symmetry 
 is broken.
There have been a lot of trials of explaining its origin, 
 and one of reliable candidates is 
 a left-right symmetric model 
 with a gauge symmetry,
 $SU(2)_L \times SU(2)_R\times U(1)_{}$\cite{Pati:1974yy, Mohapatra:1974hk, Senjanovic:1975rk}. 
Note that some grand unified theories (GUTs),
 such as $SO(10)$ GUT, 
 contains this 
 left-right symmetric gauge group. 
Anyhow, if $SU(2)_R\times U(1)_{}$ is 
 spontaneous broken to $U(1)_Y$, 
 this is an origin of breaking of left-right parity symmetry 
 in the SM. 
For this purpose, 
 we must extend a Higgs sector which contains
 new Higgs fields with quantum charges of 
 $SU(2)_R\times U(1)_{}$.   

On the other hand, 
 supersymmetry (SUSY) is one of the 
 most promising candidate beyond the SM, 
 since the existence of 
 dark matter candidate and a success of 
 gauge coupling unification.   
In the minimal set up of the supersymmetric SM (MSSM),
 the left-right symmetry can play a important role 
 to avoid R-parity violation, 
 since $U(1)_{B-L}$ symmetry (a part of left-right
 symmetric models)
 is often related to the R-charge 
 as $R=(-1)^{B-L}$\cite{Mohapatra:1986su}.
In this case,
 the $SU(2)\times U(1)_{}$ gauge symmetry is broken down to $U(1)_Y$ by the 
 triplet Higgs fields with $B-L=\pm 2$,
 and the double-charged Higgs fields are predicted.
Furthermore, it also resolves the strong CP problem 
 and the SUSY CP problem\cite{Mohapatra:1995xd, Mohapatra:1996vg}.

The SUSY gauge theory has an important feature, that is,  from
 the theoretical point of view, 
 non-perturbative effects can be calculated 
 in SUSY strong gauge theory. 
We know that the strong gauge dynamics 
 plays important role in particle physics.
The spontaneous breaking of the chiral symmetry in QCD 
 is a typical example of the spontaneous symmetry breaking.
This idea of the strong gauge dynamics applying to 
 the electroweak symmetry breaking as a scale up of QCD 
 is so-called technicolor model~\cite{Weinberg:1975gm, Eichten:1979ah}.
The electroweak scale is given by the dynamical
 scale of the technicolor (around TeV) scale, so that the scenario
 is a natural solution of the hierarchy problem.
In the case of non-SUSY theory, however,  
 it is unclear for us to
 obtain non-perturbative vacuum structure in technicolor models,
 where we usually take 
 non-trivial dynamical assumptions 
 for 
 flavor sector,  
 e.g. walking or conformal technicolor
 models~\cite{Yamawaki:1985zg, Luty:2004ye}
 or topcolor models~\cite{Hill:1991at}. 
In the framework of SUSY, 
 non-perturbative effects can be calculable in 
 a strong gauge theory. 
For example, 
 we can calculate non-perturbative effects 
 in a superpotential by using a dual description, 
 which makes us know 
 correct vacuum structure in a strong dynamics. 
The hierarchical flavor structures of this model
 may be suggested by the higher order coupling with 
 non-trivial scale dimension $d$ of 
 $\bar{\psi}{\psi}$ condensation which gives quark/lepton masses.
This value of $d$ is often communicated with R-charge
 and the dimension of the chiral operator.
For the recent work on the flavor structure of SUSY topcolor model, 
 see Refs.~\cite{Evans:2010ed, Fukushima:2010pm}.
There are some related models that provide Yukawa
 hierarchy through the anomalous dimension
 in which the MSSM sector can couple to the
 superconformal sector~\cite{Nelson:2000sn}.

If the SUSY is the underlying theory of our nature, it must be broken 
at intermediate scale between Planck scale and electroweak scale.
An idea of a dynamical SUSY breaking~\cite{Witten:1981nf} is 
 one of the most attractive scenarios which can explain 
 why the SUSY broken scale is much smaller than Planck scale. 
As for the dynamical SUSY broken models,  
 they are based on a dynamics 
 of $\mathcal{N}=1$ SQCD with various number 
 of color $N_C$ and flavor $N_F$.  
For example, 
 modified moduli space model is represented 
 in case of $N_F=N_C$~\cite{Intriligator:1996pu, Izawa:1996pk}, 
 and case of $N_F > N_C$ suggests model with 
 meta-stable vacua~\cite{Intriligator:2006dd}.

In this paper, 
 we suggest simple models of spontaneous parity violation 
 in SUSY strong gauge theory. 
We focus on left-right symmetric model and 
 investigate vacuum with spontaneous parity violation.  
Non-perturbative effects are calculable in 
 SUSY gauge theory, and 
 we suggest two new models. 
In both models, left-right symmetry breaking (and also additional $U(1)$) 
 are triggered by vacuum expectation values (VEVs) of
 $SU(2)_R$ triplets or doublets Higgs fields.

\section{A model of spontaneous parity violation}

In this section we show a model of 
 spontaneous parity violation in the SUSY 
 theory. 
We first show a basic idea, and next try to modify the model 
 by use of strong gauge dynamics. 
We will construct a model whose 
 dimensional scales are all originated from 
 the strong gauge dynamics.

\subsection{Basic structure}

We first explain the brief introduction
 to models of spontaneous parity violation. 
A model shown here is based on a gauge group $SU(2)_L \times SU(2)_R
\times U(1)_{B-L}$.   
We only focus on a Higgs sector which contains
 the following ($SU(2)_{L,R}$) doublets,
\begin{equation} \label{eq:d}
\varphi_{L}=(2,1,1), \ \ 
\tilde{\varphi_{L}}=(2,1,-1), \ \ 
\varphi_{R}=(1,2,1), \ \  
\tilde{\varphi_{R}}=(1,2,-1), \ \  
\end{equation} 
and the triplets 
\begin{equation} 
\Phi_L = (3,1,0), \ \ 
\Phi_R = (1,3,0). \ \ 
\end{equation} 
We consider a renormalizable 
 superpotential $W_{LR}$ which has left-right parity symmetry,
 $L\leftrightarrow R$, as  
\begin{equation}
W_{LR}= m(\tilde{\varphi_L} \varphi_L + \tilde{\varphi_R} \varphi_R)
+\frac{M}{2}\rm{Tr}(\Phi_L^2+\Phi_R^2)
+h(\tilde{\varphi_L} \Phi_L \varphi_L +\tilde{\varphi_R} \Phi_R \varphi_R),
\end{equation} 
where the coupling $h$ is $\mathcal{O}(1)$ and
 $m$ and $M$ have mass dimension.
The scale of $m$ and $M$ are related to the spontaneous 
 parity breaking. 
The vacuum of this model should satisfy F-flatness conditions, and 
 for the moment we ignore the gauge indexes in
 the equations, for simplicity.
The F-flatness conditions are given by 
\begin{align}\label{eq:vec}
F_{\varphi_I} = \tilde{\varphi_I}(m + h\Phi_I ), \ & \  
F_{\Phi_I} = M \Phi_I + h \tilde{\varphi_I} \varphi_I,
\end{align}
where $I= L, R$.
The first equation induces two types of the solution;  
One is a vacuum of 
 $\langle \varphi_I \rangle = \langle \varphi_I \rangle = 0$ and
 $\langle \Phi_I \rangle=0$. 
The other is a vacuum of
 $\langle \varphi_I \varphi_I \rangle = \frac{Mm}{h^2}$ and
 $\langle \Phi_I \rangle=-\frac{m}{h}$.
Remind that 
 we can always take the above 
 two solutions independently of the index $I=L, R$,  
 therefore, $SU(2)_R$ broken vacuum can be easily obtained 
 by choosing $\langle \Phi_L \rangle=0$
 and $\langle \Phi_R \rangle=-\frac{m}{h}$.\footnote{
In Ref. \cite{Babu:2008ep},  
 they induce 
 the left-right breaking vacua 
 without the neutral triplet Higgs fields,
 $\Phi, \tilde{\Phi}$, where 
 the left-right breaking vacua can be found at 
 1-loop level potential. 
}
These two mass scales of $m$ and $M$ can be taken
 large enough to satisfy the current experimental 
 bound.

This structure can be also 
 reproduced in a model with charged triplets \cite{Aulakh:1997ba}, 
 in which 
 the doublets $\varphi_I$, $\tilde{\varphi_I}$ are replaced by 
 triplet fields, 
\begin{align} 
&\Omega_L = (3,1,2), \ \ \tilde{\Omega_L} =(3,1,-2), \\
&\Omega_R = (1,3,2), \ \ \tilde{\Omega_R} =(1,3,-2), 
\end{align} 
and then the superpotential in Eq.(\ref{eq:vec}) becomes 
\begin{equation}
W_{LR}= m_\Omega \rm{Tr} (\Omega_L \tilde{\Omega_L} +\Omega_R \tilde{\Omega_R})
+\frac{M_\Phi}{2} \rm{Tr}( \Phi_L^2 +  \Phi_R^2 )
+hTr (\Omega_L \Phi_L \tilde{\Omega_L}+\Omega_R \Phi_R \tilde{\Omega_R} ).
\end{equation} 
A similar analysis can show an existence of the left-right
 asymmetric vacua also in this setup. 
One can really find the suitable left-right breaking vacua 
 from the analysis including the MSSM matter fields
 and D-flatness condition~\cite{Aulakh:1997fq}.
This triplet model has several advantages compared to the doublet model.
For example, 
 the triplet fields have parity-even charges and the
 R-parity does not break. 
It is also possible to have a tiny neutrino
 mass via see-saw mechanism through the
 triplet Higgs fields~\cite{Mohapatra:1979ia}.

Above two models contain initial mass scales (parameters),  
 which are necessary to give a parity breaking vacua and 
 independent of weak scale (but just 
 larger than the scale).  
These models are simple and complete in a sense, 
 however, we would like to consider more attractive 
 models which have no initial mass parameters. 
In the following subsections, 
 we try to modify the above
 models and achieve the spontaneous parity breaking 
 without mass parameters. 
Where 
 all dimensional scales are originated from 
 the strong gauge dynamics.

\subsection{Model improvement}

In order to avoid the ad-hoc mass scales,  
 we introduce a new gauge symmetry which becomes 
 strong at a large scale of $\Lambda$, 
 and the parity symmetry is expected to be
 broken by the
 strong SUSY dynamics.  
Also, we will consider a situation where 
 left-right Higgs sectors
 are coupled with each other 
 through Yukawa-type interactions. 

Let us introduce a new gauge dynamics 
 $SU(2)_H$, and 
 fours on the gauge symmetry  
 $SU(2)_H  \times SU(2)_L \times SU(2)_R \times U(1)_{B-L}$. 
A SUSY $SU(2)$ gauge theory with certain number of flavors
 becomes 
 strong at low energy, and has a VEV 
 of composite scalar fields through 
 a non-perturbative deformed moduli space~\cite{Seiberg:1994pq}. 
We introduce fields, $L_a^\alpha$ and $R_a^\alpha$, 
 which are charged under both 
 $SU(2)_H$ and $SU(2)_{L,R}$, 
 where $a,b=1,2$ ($\alpha=1,2$) denotes  
 $SU(2)_{L,R}$ ($SU(2)_C$) indexes. 
Under ($SU(2)_H$, $SU(2)_L$, $SU(2)_R$, $U(1)_{B-L}$), 
 they are given by 
\begin{align} 
L = (2, 2, 1,0), \ \  R = (2,1,2,0).  
\end{align} 
This field content is also used in 
 a SUSY version of the minimal
 technicolor model~\cite{Luty:2000fj}.
In order to realize spontaneous parity violation
 (and also for a cancellation of $B-L$ gauge anomaly), 
 we introduce triplet fields of 
 left-right Higgs sector
 as, 
\begin{align} 
&\Omega_L = (1,3,1,2), \ \ \tilde{\Omega_L} =(1,3,1,-2), \\
&\Omega_R = (1,1,3,2), \ \ \tilde{\Omega_R} =(1,1,3,-2), \\
&\Phi_L = (1,3,1,0), \ \ \Phi_R =(1,1,3,0). 
\end{align} 
We also introduce additional gauge singlet fields $S_L$ and $S_R$ 
which connect between strong gauge sector and left-right Higgs sector.
Then,
 a tree level 
 superpotential is given by 
\begin{align}\label{eq:tree}
W_{tree}=&
\lambda ( S_L L L +S_R R R) 
+ a 
\left\{ 
S_L \rm{Tr}[\Omega_L \tilde{\Omega_L}] + S_R \rm{Tr}[\Omega_R \tilde{\Omega_R}]
\right\}
+
b
\left\{ 
S_L \rm{Tr}\Phi_L^2 + S_R \rm{Tr}\Phi_R^2
\right\} 
\nonumber \\
&+ 
y
\left\{ 
\rm{Tr} \Omega_L \Phi_L \tilde{\Omega_L} + \rm{Tr} \Omega_R \Phi_R \tilde{\Omega_R}
\right\}
+W(S_L,S_R), 
\end{align}
where the coupling constants $\lambda$, $a$, $b$ and $y$ are 
 $\mathcal{O}(1)$ coefficients, and 
 $W(S_L,S_R)$ is the superpotential which contains
 only $S_L$ and $S_R$.
Here 
 we impose $Z_3$ discrete symmetry, whose charged are given by 
\begin{align}
\omega   &: L, S_L, \Omega_L, \tilde{\Omega_L}, \Phi_L, \\ 
\omega^2 &: R, S_R, \Omega_R, \tilde{\Omega_R},  \Phi_R, 
\end{align}
which restrict couplings such as $S_L R R$. 
We must take a setup that 
 the superpotential has no global $U(1)$ symmetry 
 which is important not to have the massless goldstone boson nor axions
 after spontaneous symmetry breaking. 
It is assumed that the global $U(1)_R$ symmetry will be broken explicitly 
 by the SUSY breaking terms. 
For the $SU(2)_H$ gauge theory, 
 it has $N_F=2$ fundamental (vector-like)
 matter fields,  
 and becomes strong 
 at low-energy scale $\Lambda$. 
Then, below the scale of $\Lambda$,  
 all $SU(2)_H$-charged matter fields
  are confined, 
 and light degrees of freedom are represented by 
 composite 
 fields as 
\begin{align}
B_L \sim \frac{L_1^\alpha L_2^\alpha \epsilon^{\alpha\beta}}{\Lambda}
,\ \
B_R \sim \frac{R_1^\alpha R_2^\alpha \epsilon^{\alpha\beta}}{\Lambda}
,\ \
\Pi =
\begin{pmatrix}
\Pi_u^0 & \Pi_d^- \\
\Pi_u^+ & \Pi_d^0
\end{pmatrix}
\sim \frac{L^\alpha R^\beta \epsilon^{\alpha\beta} }{\Lambda},
\nonumber 
\end{align}
where $\Pi$ field is a bi-doublet under the $SU(2)_L$ and $SU(2)_R$.
Due to the strong dynamics of $SU(2)_H$, 
 a quantum constraint 
 is given by 
\begin{align}
\det{\Pi}-B_L B_R=\Pi^0_u \Pi^0_d -\Pi^+_u \Pi^-_d -B_L B_R=f^2, 
\nonumber 
\end{align}
where $f = \Lambda/4\pi$.
If we set this scale $\Lambda \sim 1$ TeV,
 it is shown that 
 $f \sim 100$ GeV.
Thus, 
 the scale of electroweak
 symmetry breaking can be related to 
 the strong $SU(2)_H$ dynamics, 
 which will be discussed later
 after including the electroweak Higgs fields. 
Anyhow, a low energy effective theory below a TeV scale
 can be described by 
 these composite fields. 
We apply a naive dimensional analysis~\cite{Manohar:1983md, Cohen:1997rt} 
 for these fields and canonical normalization for
 all the composite fields,
 we can obtain the following 
 effective superpotential, 
\begin{align}
W_{\rm{eff}} =&
\lambda f (S_L B_L + S_R B_R) 
+ a 
\left\{ 
S_L \rm{Tr}[\Omega_L \tilde{\Omega_L}] + S_R \rm{Tr}[\Omega_R\tilde{\Omega_R}]
\right\}
+
b
\left\{ 
S_L \rm{Tr}\Phi_L^2 + S_R \rm{Tr}\Phi_R^2
\right\} 
\nonumber \\
&+
y
\left\{ 
\rm{Tr} \Omega_L \Phi_L \tilde{\Omega_L} + \rm{Tr} \Omega_R \Phi_R \tilde{\Omega_R}
\right\}
+W(S_L, S_R) \nonumber \\ 
&+ X(\Pi_u^0 \Pi_d^0-\Pi_d^+ \Pi_d^- -B_L B_R -f^2 ),
\end{align}
where $X$ denotes the Lagrange multiplier to constrain the 
quantum deformed moduli spaces.

The supersymmetric vacua is given by the following F-flatness conditions 
\begin{align}
\frac{\partial W}{\partial S_L} 
&=  \lambda f B_L + a\rm{Tr}[\Omega_L\tilde{\Omega_L}]
+b  \rm{Tr}[\Phi_L^2] + \frac{\partial W(S_L, S_R)}{\partial S_L}=0,
 \nonumber  \\
\frac{\partial W}{\partial S_R}
&=  \lambda f B_R + a\rm{Tr}[\Omega_R\tilde{\Omega_R}]
+b \rm{Tr}[\Phi_R^2] + \frac{\partial W(S_L, S_R)}{\partial S_R} =0,  
 \nonumber  \\
\frac{\partial W}{\partial \Pi_{u,d}^0}
&=  X \Pi_{d,u}^0 = 0, \ \  
\frac{\partial W}{\partial \Pi^\pm}
=  -X \Pi^\mp  =0, 
 \nonumber  \\ 
\frac{\partial W}{\partial \omega_L^0}
&=  \tilde{\omega}_L^0 (a S_L + \frac{y \delta_L}{\sqrt{2}}) =0,  \ \ 
\frac{\partial W}{\partial \omega_R^0}
=   \tilde{\omega}_R^0 (a S_R +\frac{y \delta_R}{\sqrt{2}} ) =0, \label{eq:LR} \\
\frac{\partial W}{\partial \tilde{\omega_L}^0}
&=   \omega_L^0 (a S_L  +\frac{y \delta_L}{\sqrt{2}} )
 =0,  \ \  
\frac{\partial W}{\partial \tilde{\omega_R}^0}
=  \omega_R^0 (a S_R + \frac{y \delta_R}{\sqrt{2}} )=0, 
 \nonumber  \\
\frac{\partial W}{\partial \delta_L^0}
&=  2 b S_L \delta_L^0 + \frac{y \omega_L^0 \tilde{\omega}_L^0}{\sqrt{2}}  =0, \ \ 
\frac{\partial W}{\partial \delta_R^0}
=  2 b S_R \delta_R^0 + \frac{y \omega_R^0 \tilde{\omega}_R^0}{\sqrt{2}}  =0,  
 \nonumber  \\ 
\frac{\partial W}{\partial B_L}
&=  \lambda f S_L - B_R X =0, \ \ 
\frac{\partial W}{\partial B_R}
=   \lambda f S_R - B_L X =0, \nonumber 
\end{align}
with quantum constraint,
 $\Pi_u^0 \Pi_d^0-\Pi^+ \Pi^--B_L B_R=f^2$. 
Here, 
 we have used the following notation for 
 the triplet fields as
\begin{align}
\Omega_L =
\begin{pmatrix}
\frac{\omega_L^+}{\sqrt{2}} & \omega_L^{++} \\
\omega_L^0 & -\frac{\omega_L^+}{\sqrt{2}} \\
\end{pmatrix}, \ \ \ 
\Omega_R =
\begin{pmatrix}
\frac{\omega_R^-}{\sqrt{2}} & \omega_R^{0} \\
\omega_R^{--} & -\frac{\omega_R^-}{\sqrt{2}} \\
\end{pmatrix}, \ \ \ 
\Phi_L =
\begin{pmatrix}
\frac{\delta_L^0}{\sqrt{2}} & \delta_L^+ \\
\delta_L^- & -\frac{\delta_L^0}{\sqrt{2}} \\
\end{pmatrix},  
\end{align}
in which superscripts show the
electro-magnetic charge, 
 $Q= T_L+T_R+{(B-L)/2}$ 
 and the field contents for $\tilde{\Omega}_L$, $\tilde{\Omega}_R$ and $\Phi_R$ 
 are understood as well.

We have assumed that charged fields do not take non-zero magnitudes of 
 VEVs, and thus 
 have dropped the charged fields, for simplicity. 
We are interested in the left-right asymmetric vacua
 which are given by a solution of  
 $\langle \omega_L^0 \rangle, \langle \tilde{\omega}_L^0 \rangle=0$ and 
 $\langle \omega_R^0 \rangle, \langle \tilde{\omega}_R^0 \rangle \ne 0 $. 
Through the equation $F_\Pi=0$, 
 we can show that 
 the electroweak symmetry is not broken ($\langle \Pi \rangle=0$),  
 if $\langle X \rangle \ne 0$.  
Thus, 
 in order to avoid the unwanted
 large scale electroweak symmetry breaking, 
 we take $\langle X \rangle \ne 0$ here, and 
 we will consider the electroweak symmetry breaking later.

As shown in Eq.(\ref{eq:LR}),  
 one can obtain the similar structure given in 
 Eq.(\ref{eq:vec}).
In this case, from the equation
 of $F_{\omega}=0$ or $F_{\omega^c}=0$, 
 the solutions for left-right Higgs 
 sectors are classified into two types of solutions, 
 that is, 
 $\langle \omega_{L, R}^0 \rangle =0$ or $\neq 0$. 
For example, if we have a solution $\langle \omega_R^0 \rangle = 0$, 
 we obtain  $\langle \delta_R^0 \rangle =0$, where  
 the $SU(2)_R$ symmetry 
 is not broken. 
On the other hand, if we have another solution of 
 $\langle \delta_R^0 \rangle = -\frac{\sqrt{2}a \langle S_R \rangle}{y}$,
 we obtain the VEVs as 
 $\langle \omega_R^0 \tilde{\omega}_R^0 \rangle =\frac{2 a b \langle S_R \rangle^2 }{y^2}$.
This is the suitable symmetry breaking of 
 $SU(2)_R \times U(1)_{B-L} \rightarrow U(1)_Y$. 
Thus, 
 as long as $\langle S_R \rangle \ne 0$ 
these two types of the solutions have different structure.
In this case, 
 we can take a different vacuum structure for $L$ and $R$
 sector as, 
 $\langle \delta_L^0 \rangle =0$ and 
 $\langle \delta_R^0 \rangle=b \langle S_R \rangle/y$.  
Where, VEVs of other fields are given by
\begin{align} 
&
\langle B_L \rangle = \lambda f \frac{ \langle S_R \rangle}{ \langle X\rangle}, \ \ \ 
\langle B_R \rangle = \lambda f  \frac{\langle S_L \rangle}{ \langle X\rangle}, \label{eq:B} \\
& 
\langle \omega_L^0 \rangle =\langle \tilde{\omega}_L^0 \rangle =0,  \ \ \ 
\langle \omega_R^0 \tilde{\omega}_R^0 \rangle =
 \frac{2ab}{y^2}\langle S_R \rangle^2. \nonumber
\end{align} 
And 
 VEVs of $B_L$, $B_R$ can be determined by solving equations of 
\begin{eqnarray}\label{eq:elr}
\frac{\partial W}{\partial S_L} &=&
\lambda f B_L + \frac{\partial W(S_L, S_R)}{\partial S_L } =0, \\ 
\frac{\partial W}{\partial S_R} &=&
\lambda f B_R + a \omega_R^0 \tilde{\omega}_R^0 +b \left(\delta_R^0\right)^2
+\frac{\partial W(S_L, S_R)}{\partial S_R } =0. \nonumber
\end{eqnarray}
The values of these VEVs fully depend on
 the details of the structure of $W(S_L,S_R)$.
Here we consider following renormalizable left-right symmetric
 superpotential, 
\begin{eqnarray}\label{eq:wlr}
W(S_L, S_R) = h (S_L^3+S_R^3),
\end{eqnarray}
where $h$ is a $O(1)$ coefficient.
The VEVs obtained by solving Eq.(\ref{eq:elr}) with
 Eq.(\ref{eq:wlr})
 are shown as 
\begin{eqnarray}
|\langle B_L \rangle | &=& \left( \frac{h y^2}{hy^2+2a^2b} \right)^{1/6}
 f,  \nonumber \\ 
|\langle B_R \rangle | &=& \left( \frac{hy^2+2a^2b}{h y^2} \right)^{1/6} f,  \\ 
|\langle X \rangle| &=&  
\frac{\lambda^{3/2}} {(3hc)^{1/4}} f. \nonumber
\end{eqnarray}
Notice that 
 we have only one energy scale $f$,
 which can appear as SUSY dynamics and its value can be taken 
 arbitrary.  
Thus, taking larger energy scale of $f$ (than the electroweak symmetry
 breaking),  
 it is possible
 to have a large scale spontaneous
 left-right parity violation.   
The order estimation gives the scale of VEVs as
$\langle S_L \rangle \sim \langle S_R \rangle \sim \langle \tilde{\delta_R} \rangle \sim f$,  
$\langle {\omega_R} \tilde{\omega_R} \rangle \sim f^2$, and
$\langle B_L \rangle \sim \langle B_R \rangle \sim f$. 
The $SU(2)_R$ symmetry is broken down to $U(1)_R$
 by $\langle \delta_R^0  \rangle$, 
 and further breakdown of the symmetry
 $U(1)_R \times U(1)_{B-L} \to U(1)_Y$ by the VEVs of 
 $\omega_R\tilde{\omega_R}$. 
Therefore, these vacua can derive left-right asymmetric solution 
 as in Ref.~\cite{Aulakh:1997fq}.
 It should be noted that,
 in this model, left and right Higgs sectors are not independent of
 each other 
 due to the quantum constraint, 
 so the asymmetric solution (vacua) is non-trivial.

Now let us investigate 
 mass spectra  
 of the Higgs supermultiplet. 
In the model, two symmetry breaking scales of $SU(2)_R$ and $U(1)_{B-L}$ 
 are the same with the VEVs of $S_L$ and $S_R$.
So all the triplet mass scale may be given by $f$. 
Similarly the masses of the singlets $B_{L, R}$ and $S_{L, R}$ are same scale $f$.
As for the composite Higgs fields $\Pi$, 
 they can have effective mass of order $f$ with Higgs bi-doublet 
 after including the electroweak Higgs field (see Eq.(\ref{22})).

Here we comment on some topics. 
The first is about field content. 
While we construct the model using the triplet fields, 
 we can also apply these mechanism 
 in the doublet fields as shown in the previous section. 
The next is about 
 the electroweak symmetry breaking.
At the present stage, 
 we do not add an elementary Higgs fields which should
 have VEVs for the 
 electroweak symmetry breaking and inducing quark/lepton masses.  
As shown in Ref.~\cite{Choi:1999yaa}, 
 the electroweak Higgs field can couple 
 to the $\Pi$ which gives rise to a SUSY mass term for Higgs fields. 
This may be a alternative solution to so-called $\mu$-problem 
 as shown next subsection (see, Eq.(\ref{22})). 
For the realistic model, soft SUSY breaking terms should be  
 included in the model and can lead to the electroweak symmetry
 breaking vacua as the usual MSSM. %
The soft masses $m_{\rm{soft}} $ of order a few hundred GeV 
 are small enough comparing to the dynamical scale $f$, 
 so that SUSY breaking effects can be treated 
 as a perturbation based on the naive dimensional analysis.  
Thus, it is possible to analyse the spectra 
 including the soft SUSY breaking terms,  
 and this situation 
 leads entirely different structure from Ref.~\cite{Choi:1999yaa}.

It is also interesting to consider the dynamical electroweak symmetry breaking 
due to this SUSY strong gauge dynamics. 
In this setup, there are the composite electroweak Higgs fields $\Pi$,
 and its VEVs itself can break the electroweak symmetry breaking. 
It is the same as the scenario of the technicolor models
 with SUSY extension~\cite{Luty:2000fj}.
Naively, we can consider a setup that
 these fields can directly couple to the elementary Higgs fields, 
 and the quark/lepton mass matrices and mixing can be obtain
 through the ordinary Yukawa couplings without 
 dangerous FCNC processes.
These mechanisms can be achieved by taking the different vacuum.
In the next section,
 we discuss the possible electroweak symmetry
 breaking in above models.

\subsection{Electroweak symmetry breaking }

We include the Higgs bi-doublets field $H$ and an additional singlet $S$.
The superpotential is given by 
\begin{eqnarray}
W_H= \tilde{h_y} LRH+\alpha S H^2+\kappa S^3.
\end{eqnarray}
Then, below the dynamical scale $\Lambda$, 
the effective superpotenial becomes 
\begin{eqnarray}
\label{22}
W_{\rm{eff}}= 
h_y f \Pi H + \alpha SH^2+\kappa S^3.
\end{eqnarray}
The F-flatness conditions for $H$, $\Pi$, and $S$ are given by 
\begin{eqnarray}
\frac{\partial W}{\partial H} &=& f h_y \Pi + \alpha S H, \ \ \
\frac{\partial W}{\partial \Pi} = X \Pi + f h_y H, \\ 
\frac{\partial W}{\partial S} &=& \alpha H^2 + 3 \kappa S^2. \nonumber 
\end{eqnarray}
Assuming $\Pi^\pm=0$, 
 the VEVs of the fields $S$ and $\Pi$ are given
 by the following relations, 
\begin{eqnarray}\label{eq:Pi}
\langle \Pi_u^0 \rangle \langle \Pi_d^0 \rangle 
&=& -\frac{3\kappa}{\alpha^3}
\frac{(fh_y)^6}{\langle X \rangle^4}, \ \ \ 
\langle S \rangle 
= \frac{(fh_y)^2}{\alpha \langle X \rangle}.
\end{eqnarray}
The remaining equations are given by
\begin{eqnarray}\label{eq:elr2}
\frac{\partial W}{\partial S_L} &=&
\lambda f B_L + \frac{\partial W(S_L, S_R)}{\partial S_L } =0, \\ 
\frac{\partial W}{\partial S_R} &=&
\lambda f B_R + a \omega_R^0 \tilde{\omega}_R^0 +b \left( \delta_R^0 \right)^2
+\frac{\partial W(S_L, S_R)}{\partial S_R } =0, 
\end{eqnarray}
with quantum constraint 
\begin{eqnarray}
\Pi_u^0 \Pi_d^0 - B_L B_R =-f^2.
\end{eqnarray}

Now let us consider the superpotential $W(S_L, S_R)$ as
\begin{eqnarray}
W(S_L, S_R) = M S_L S_R +h (S_L^3+S_R^3),
\end{eqnarray}
where we introduce the vector-like mass term $M S_L S_R$
 for the singlet fields. 
We here introduce additional mass 
 scale of $M$ as the singlet mass,
 since it is very difficult to induce two 
 different scales, left-right symmetry breaking 
 and electroweak scales, only from $f\ (\Lambda)$. 
Although it is contradict our motivation, 
 ``all scale must be introduced dynamically'', 
 it might be a minimum setup of reproducing 
 suitable scales of left-right symmetry breaking 
 and electroweak scales. 
We will soon know they are suitably induced from 
 $M$ and $f\ (\Lambda)$. 
 
Actually this new parameter (scale) $M$ can be regarded as
 the SUSY breaking effects. 
Since, without the term $M S_L S_R$,
 the superpotential in Eq.~(\ref{eq:tree})  
 has a global $U(1)_R$ symmetry where both 
 the $L$- and $R$-subscripted fields have its charge $2/3$.
Thus it could be natural to assume the $U(1)_R$ symmetry is broken 
  by the soft SUSY breaking terms. 
For example, 
 an introduction of 
 a spurion field $\xi=F_\xi \theta^2$ with $U(1)_R$-charge of 2  
 can induce the term $M S_L S_R$ through 
 $D$-term interaction  
 $[\xi^\dagger S_L S_R]_D$ as well as 
 SUSY breaking terms.

Assuming the hierarchy between $f$ and $M$, 
 which is reasonable assumption and makes us easily see the vacuum
 structure, 
 we find three different types of vacuum
 solutions; 
(i): $\langle B_L \rangle \sim \langle B_R \rangle \sim M^2/f$ and 
 $\langle X \rangle \sim f^2/M$, 
(ii): $\langle B_L \rangle = \langle B_R \rangle = 0$ and 
  $\langle X \rangle \sim f$,    
(iii): $\langle B_L \rangle \sim \langle B_R \rangle = f$ and  
 $\langle X \rangle \sim M$. 
The cases of (i) and (ii) are
 out of our interest.
It is because, 
 in case of (i), the triplet Higgs fields get the mass from the VEVs of 
 $\langle S_{L, R} \rangle \sim \langle B_{R, L} X \rangle/f \sim M$, 
 so that the VEVs of $\Pi$ are estimated as
 $\langle \Pi \rangle \sim M^2/f$ from Eq.~(\ref{eq:Pi}), 
 which means the VEV of $\Pi$ is too large 
 since we assume $M \gg f$. 
As for the case of (ii), obviously the vacua do not suggest
 the left-right symmetry breaking.
Thus, let us take the case of (iii).

From Eqs.(\ref{eq:B}),  the left-right breaking scale is given the 
 scale of $M$,
 while the VEVs of $S$ and $H$ are of order $f^2/M$. 
Thus we can easily obtain the hierarchy between 
the left-right breaking scale and electroweak scale.
Furthermore, the composite Higgs VEVs 
 is of order $\langle \Pi \rangle \sim \frac{f^3}{M^2}$. 
Taking 
 a numerical example as 
 $a=1, b=-1/3, \lambda=y=h=h_y=1, \kappa=-1/3$, 
 $f= \frac{\Lambda}{4\pi}=1.5\ \rm{TeV}$, and $M=15\ \rm{TeV},$
 the VEVs for the case (iii) are given by  
\begin{eqnarray}
&& |\langle B_L \rangle| \sim |\langle B_R \rangle| \sim 1.5 {\rm [TeV]}, \ \ 
 |\langle X \rangle| \sim 15 {\rm [TeV]},  
\end{eqnarray}
This vacuum has correct 
 electroweak symmetry breaking vacua with large scale 
 left-right symmetry breaking. 
It is because the electroweak symmetry breaking vacua is dominated by 
 $\langle H\rangle \sim f^2/M$, and the left-right
 symmetry scale is given by 
 the mass scale of $M$.

Here we briefly show the mass spectra in the supersymmetric model.
As already explained, 
 the charged triplet Higgs masses $\omega_L^{++}$ $\omega_R^{++}$
 are given by the VEVs of $S_{L, R} \sim M$.
On the other hand, 
 the neutral components $\delta$ are mixed with $S_{L, R}$ 
 or composite fields $B_{L, R}$, 
 where their VEVs are $M$, $M$ and $f$, respectively.
As we expected, they have a TeV scale mass.
For the 
 neutral components of
 doublet Higgs $H^0_u$ and $H^0_d$, 
 they can mix with 
 $\Pi^0_u$, $\Pi^0_d$, and $S$.
And,  
 the mass matrix of these fields $(S, H, \Pi)$ are given 
\begin{eqnarray}\label{eq:mass}
\begin{pmatrix}
\langle S \rangle & \langle H \rangle &0 \\
\langle H \rangle &\langle S \rangle & f \\
0 & f & 0 \\
\end{pmatrix}
\end{eqnarray}
The mass eigenvalues of this mass matrix are roughly
 estimated by $(f, f, f/M^2)$.
Thus, there exists possibly one light Higgs component in the model. 

For the realistic electroweak symmetry breaking,
 the following  
 Fermi relation should be satisfied, 
\begin{eqnarray}
(176)^2 \rm{GeV}^2 = \langle \Pi_u^o \rangle^2 + \langle \Pi_d^0 \rangle^2 
+ \langle H_u^0 \rangle^2 +\langle H_d^0 \rangle^2.
\end{eqnarray}
Notice that these two symmetry breakings occur 
 by the strong gauge dynamics in both cases 
 with and without SUSY breaking effect. 
We can expect the 
 realistic electroweak symmetry breaking vacua 
 is obtained even 
 after additional soft SUSY breaking terms are induced, 
 and its reason is shown just below. 
As for oblique electroweak contributions~\cite{Peskin:1990zt},  
 the dynamical breaking models can give
 a certain value of the electroweak S-parameter in general, 
 which is strongly constrained in conventional technicolor models.
In our model,
 SUSY strong dynamics also give a contribution to it. 
However the leading order contribution 
 is small comparing to previous technicolor models.
It is because the electroweak symmetry breaking scale is
 given by $f^2/M$, and which means 
 there is further suppression in the case of $f \sim 1$[TeV]. 
This naive dimensional analysis can also apply to
 the soft SUSY breaking terms. 
This is the reason why desirable vacuum is expected to be obtained 
 after including SUSY breaking effects. %

\section{Model II (IR free model)}

In the previous section, we considered the strongly coupled 
 gauge dynamics and light degrees are confinement. 
Next, we consider another example for spontaneous parity violation   
 based on the doublet model. 
We use the fact that strong SUSY dynamics which has various vacuum
 structures can exhibit 
 meta-stable SUSY breaking\cite{Intriligator:2006dd}.
The simplest way to construct a model
 with spontaneous parity violation is to consider both 
 the left and right gauge sectors individually as 
 the model explained in Eq.~(\ref{eq:d}),  
 and 
 combine these two sectors with different vacuum alignment. 
Let us consider the gauge group
 $SU(3)_{H_I}  \times SU(2)_I \times U(1)_{B-L}$, where 
 the gauge $SU(3)_{H_I}$ is additional strong interaction and 
 $I$ means L(eft) or R(ight).
The chiral multiplets $Q^a$ and $\tilde{Q}^a$, (a=1, 2)
 have following quantum numbers 
\begin{eqnarray}
Q^a \sim (3, 2, 1),  \ \ \ 
\tilde{Q}^a \sim (\bar{3}, 2, -1),
\end{eqnarray}
and a common vector-like mass $m$.
This theory is $N_C=3$  and $N_F=4$ in SQCD, 
 and has a magnetic dual description valid at the low
 energy scale\cite{Seiberg:1994pq}. 
Below the scale of $m \ll \Lambda$,
 the magnetic dual theory has dual matter fields
 $\varphi^a$, $\tilde{\varphi}^a$ 
 and meson fields $\Phi^{ab}$ with the quantum number 
\begin{eqnarray}
\varphi^a \sim (2, -1),  \ \
\tilde{\varphi}^a \sim (2, -1), \ \
\Phi^{ab} \sim {\bf 1+ \rm{Adj.} }.
\end{eqnarray}
In the case the magnetic dual fields are equivalent to 
the s-confinement fields given by the mesons and baryons.  

The general dual superpotential is given by 
\begin{eqnarray}\label{eq:f}
W= h \rm{Tr} (\tilde{\varphi} \Phi \varphi)
 + \alpha \Lambda m\ \rm{tr}(\Phi)+\frac{c}{\Lambda} \rm{det}\Phi, 
\end{eqnarray}
where the couplings, $h$, $\alpha$,
 and $c$ are $\mathcal{O}(1)$ parameters, and $\rm{tr}$ means 
 the trace over gauge and flavor indexes.
If we neglect the third term in Eq.(\ref{eq:f}),  
 where the scale $\Lambda$ is much larger than 
 the VEVs of $\Phi$,
 SUSY is broken 
 by the rank conditions~\cite{Intriligator:2006dd}.
This SUSY breaking vacua is regarded as meta-stable vacua, 
 since  
 there is a SUSY vacuum in Eq.(\ref{eq:f}).

We see more details of this model.
At first, let us consider a effective superpotential
\begin{eqnarray}
W= h \rm{Tr} (\tilde{\varphi} \Phi \varphi)
 + \alpha \Lambda m\ \rm{tr}(\Phi). 
\end{eqnarray}
Here we denoted these doublets $\varphi$ and meson $\Phi$  as
\begin{eqnarray}
(\varphi^T)^a = (\varphi^0 ,  \varphi^- )^a , \ \ \ 
\Phi^{ab} = 
\begin{pmatrix}
\frac{S + \delta^0}{\sqrt{2}} & \delta^+ \\
\delta^- &  \frac{S - \delta^0}{\sqrt{2}} 
\end{pmatrix}^{ab},
\end{eqnarray}
where superscript shows the electromagnetic charge of
 $SU(2)$ doublet and triplet, 
 and the field $S$ is 
 gauge singlet field.
It is always possible to take the VEVs of
 $\varphi$ as 
 $\langle  \varphi^T \rangle^a=(\langle \varphi^0 \rangle, 0 )^a $
 by using the degrees of freedom of the 
 gauge and flavor transformations.
Then, 
 the second term in the superpotential
 $\alpha \Lambda m \rm{tr} (\Phi)$ is expressed 
 as 
\begin{eqnarray}
\alpha \Lambda m \rm{tr} (\Phi)
 = \sqrt{2}\alpha \Lambda m (S^{11}+S^{22}),
\end{eqnarray}
and this mass parameter preserves the $SU(2)$ gauge symmetry but breaks the flavor $SU(2)$ symmetries
to its diagonal form. 
The F-flatness conditions for $S$ and $\delta^0$ are given by
\begin{eqnarray}
F_{S^{11, 22}}&=&\sqrt{2} \alpha \Lambda m + \frac{(\tilde{\varphi^0}{\varphi^0})^{11, 22}}{\sqrt{2}}=0, \nonumber \\
F_{(\delta^0)^{ab}}&=& \frac{(\tilde{\varphi^0}{\varphi^0})^{ab}}{\sqrt{2}}=0. \nonumber 
\end{eqnarray}
{}From these equations, the minimum is given by 
\begin{eqnarray}
\langle \tilde{\varphi^0 }\varphi^0 \rangle^{ab}
 =-\frac{\alpha \Lambda m}{2} \ne 0. 
\end{eqnarray}
The above solutions mean that 
 gauge group $SU(2) \times U(1)$ reduces to $U(1)$ 
 as well as SUSY is broken in the vacuum.

Next, we find the true vacuum with 
 the third term of Eq.(\ref{eq:f}). 
The true vacua can be obtained by solving
 the full superpotential in Eq.(\ref{eq:f}).
We can find the supersymmetric vacuum at  
\begin{eqnarray}
\langle S^{11} \rangle \sim \langle S^{22} \rangle  \sim \left(  \Lambda^2 m \right)^{1/3} ,
\end{eqnarray}
and doublet fields can not have their VEVs. 
This singlet VEV gives masses of the vector-like Higgs doublets.
This leads to the SUSY 
 solution, and the gauge group $SU(2) \times U(1)$ never breaks 
 into its diagonal form.

Finally, let us consider the larger gauge group as
$SU(3)_{H_L} \times SU(2)_L \times SU(3)_{H_R} 
\times SU(2)_R \times U(1)_{B-L}$ with left-right symmetry, 
 and  
 combine above two results. 
We take a vacuum where 
 the meta-stable vacua for right-sector, and 
 the SUSY true vacua for left-sector. 
Then, 
 we obtain parity breaking vacuum. 
 Please note that  since the full gauge theory of this model is 
 chiral gauge model so our results do not 
 conflict the Vafa-Witten theorem~\cite{Vafa:1983tf}.
 
The flavor structure of these models are expected to
 be obtained by introducing
 elementary Higgs fields $H$ through the 
 Yukawa couplings. 
Anyhow, notice again, 
 the electroweak symmetry breaking and 
 SUSY breaking are triggered simultaneously 
 in this model.  
We can expect that the suitable soft SUSY breaking
 terms are induced from the right-sector.

\section{Summary and discussions}

We have suggested simple models of spontaneous parity violation 
 in SUSY strong gauge theory. 
We have focused on left-right symmetric model and 
 investigate vacuum with spontaneous parity violation.  
Non-perturbative effects are calculable in 
 SUSY gauge theory, and 
 we have suggested two new models. 
The first model showed confinement, 
 and the second model had a dual description 
 of the theory. 
The left-right symmetry breaking and 
 electroweak symmetry breaking are 
 simultaneously occurred 
 with the suitable energy scale hierarchy. 
The second model had also induced spontaneous 
 SUSY breaking. 

In detail, 
 the first model suggested that left-right sectors couple 
 to new strong gauge fields 
 which are confined at low energy. 
Scale of left-right symmetry breaking is determined 
 by the dynamical scale of new strong gauge theory, where 
 we consider a $\mathcal{N}=1$ SQCD with $N_C=2$ and $N_F=4$. 
Our setup is related to the SUSY technicolor model, 
 so it is easy to make the model couple to the electroweak Higgs. 
We can realize 
 the left-right symmetry breaking and the electroweak symmetry breaking 
 simultaneously with the suitable energy scale hierarchy. 
This structure has several advantages compared to the MSSM. 
The scale of the Higgs mass scale (left-right breaking scale)
 and that of VEVs are
 different,
 so the SUSY little hierarchy problems are absent. 
At same time, the SUSY breaking terms, which are expected to
 be smaller than dynamical scale, can be treated 
 perturbatively by assuming the canonical Kahler potential
 for all the composite fields.

The second model also has the structure,  
 that is, left-right Higgs fields couple to
 new strong gauge fields.
The dual description of the model is possible,
 and one can find the $SU(2)_R$ broken vacua of meta-stable, 
 which means  
 this model also induces spontaneous SUSY breaking. 
The UV completion of this setup only includes
 vector-like left-right two Higgs fields, 
 therefore this setup is a minimal for left-right symmetric scenario.  
The low energy effective theory of this model
 should be the MSSM with induced soft SUSY breaking terms, and 
 all the spectra are calculable.

Finally, 
 we comment on a recent Tevatron experiment~\cite{Aaltonen:2011mk} which 
 reported the anomalous excess in $W_{jj}$ dijet events and 
 implied the existence of the resonance state 
 with mass $M \sim 150$[GeV]. 
Their simple explanation can be given by 
 new particle originated from new strong interaction like 
 a technicolor model~\cite{Eichten:2011sh}.
It is naively expected that there exists
 a possible particle candidate in our model. 
For example, in our first model, a new particle from 
 the mixing of $\Pi$ and $H$ in 
 Eq.(2.31) could have a mass around $150$[GeV].


\vspace{1cm}

{\large \bf Acknowledgments}\\

\noindent
We acknowledge H. Terao for fruitful discussions and comments.
This work is partially supported by Scientific Grant by Ministry of 
 Education and Science, Nos. 20540272, 20039006, and 20025004. 
The work of HO is supported by the JSPS Grant-in-Aid for Scientific Research (S) No. 22224003.


\end{document}